\PassOptionsToPackage{unicode}{hyperref}
\PassOptionsToPackage{hyphens}{url}
\PassOptionsToPackage{dvipsnames,svgnames,x11names}{xcolor}
\documentclass[
]{article}

\usepackage{amsmath,amssymb}
\usepackage{lmodern}
\usepackage{iftex}
\ifPDFTeX
  \usepackage[T1]{fontenc}
  \usepackage[utf8]{inputenc}
  \usepackage{textcomp} % provide euro and other symbols
\else % if luatex or xetex
  \usepackage{unicode-math}
  \defaultfontfeatures{Scale=MatchLowercase}
  \defaultfontfeatures[\rmfamily]{Ligatures=TeX,Scale=1}
\fi
\IfFileExists{upquote.sty}{\usepackage{upquote}}{}
\IfFileExists{microtype.sty}{% use microtype if available
  \usepackage[]{microtype}
  \UseMicrotypeSet[protrusion]{basicmath} % disable protrusion for tt fonts
}{}
\usepackage{xcolor}
\ifx\paragraph\undefined\else
  \let\oldparagraph\paragraph
  \renewcommand{\paragraph}[1]{\oldparagraph{#1}\mbox{}}
\fi
\ifx\subparagraph\undefined\else
  \let\oldsubparagraph\subparagraph
  \renewcommand{\subparagraph}[1]{\oldsubparagraph{#1}\mbox{}}
\fi

\usepackage{longtable,booktabs,array}
\usepackage{calc} % for calculating minipage widths
\usepackage{etoolbox}
\makeatletter
\patchcmd\longtable{\par}{\if@noskipsec\mbox{}\fi\par}{}{}
\makeatother
\IfFileExists{footnotehyper.sty}{\usepackage{footnotehyper}}{\usepackage{footnote}}
\makesavenoteenv{longtable}
\usepackage{graphicx}
\makeatletter
\def\maxwidth{\ifdim\Gin@nat@width>\linewidth\linewidth\else\Gin@nat@width\fi}
\def\maxheight{\ifdim\Gin@nat@height>\textheight\textheight\else\Gin@nat@height\fi}
\makeatother
\setkeys{Gin}{width=\maxwidth,height=\maxheight,keepaspectratio}
\makeatletter
\def\fps@figure{htbp}
\makeatother
\newlength{\cslhangindent}
\newlength{\csllabelwidth}
\newlength{\cslentryspacingunit} % times entry-spacing
 {% don't indent paragraphs
  \setlength{\parindent}{0pt}
  \ifodd #1
  \let\oldpar\par
  \def\par{\hangindent=\cslhangindent\oldpar}
  \fi
  \setlength{\parskip}{#2\cslentryspacingunit}
 }%
 {}
\usepackage{calc}

\usepackage[noblocks]{authblk}

\usepackage{mathtools}
\usepackage[left]{lineno}
\usepackage{tabularx}
\usepackage{longtable}
\usepackage{multirow}
\usepackage{setspace}
\renewcommand{\abstractname}{Summary}
\usepackage{bm}
\usepackage{rotating}
\usepackage{natbib}
\usepackage{tabularx}
\usepackage[colorlinks=false]{hyperref}
\usepackage{longtable}
\usepackage[nointegrals]{wasysym}
\usepackage{svg}
\usepackage{amsmath}
\usepackage{booktabs}
\usepackage[tableposition=bottom]{caption} % Spaces the caption properly
\usepackage{rotating}
\usepackage{subfig}
\usepackage[a4paper, total={6in, 8in}]{geometry}
\usepackage{xhfill}

\makeatletter
\@ifpackageloaded{caption}{}{\usepackage{caption}}
\AtBeginDocument{%
\ifdefined\contentsname
  \renewcommand*\contentsname{Table of contents}
\else
  \newcommand\contentsname{Table of contents}
\fi
\ifdefined\listfigurename
  \renewcommand*\listfigurename{List of Figures}
\else
  \newcommand\listfigurename{List of Figures}
\fi
\ifdefined\listtablename
  \renewcommand*\listtablename{List of Tables}
\else
  \newcommand\listtablename{List of Tables}
\fi
\ifdefined\figurename
  \renewcommand*\figurename{Figure}
\else
  \newcommand\figurename{Figure}
\fi
\ifdefined\tablename
  \renewcommand*\tablename{Table}
\else
  \newcommand\tablename{Table}
\fi
}
\@ifpackageloaded{float}{}{\usepackage{float}}
\floatstyle{ruled}
\@ifundefined{c@chapter}{\newfloat{codelisting}{h}{lop}}{\newfloat{codelisting}{h}{lop}[chapter]}
\floatname{codelisting}{Listing}

\makeatother
\makeatletter
\@ifpackageloaded{caption}{}{\usepackage{caption}}
\makeatother
\makeatletter
\@ifpackageloaded{tcolorbox}{}{\usepackage[many]{tcolorbox}}
\makeatother
\makeatletter
\@ifundefined{shadecolor}{\definecolor{shadecolor}{rgb}{.97, .97, .97}}
\makeatother
\ifLuaTeX
  \usepackage{selnolig}  % disable illegal ligatures
\fi
\IfFileExists{bookmark.sty}{\usepackage{bookmark}}{\usepackage{hyperref}}
\IfFileExists{xurl.sty}{\usepackage{xurl}}{} % add URL line breaks if available
\hypersetup{
  pdftitle={A sequential Monte Carlo algorithm for data assimilation problems},
  pdfauthor={Kwaku Peprah Adjei; Robert B. O'Hara; Nick Isaac; Diana Bowler; Rob Cooke},
  colorlinks=true,
  linkcolor={blue},
  filecolor={Maroon},
  citecolor={blue},
  urlcolor={blue},
  pdfcreator={LaTeX via pandoc}}

\title{A sequential Monte Carlo algorithm for data assimilation problems in ecology}

\author[1,2]{Kwaku Peprah Adjei}
\author[3]{Rob Cooke}
\author[3]{Nick Isaac}
\author[1,2]{Robert B. O'Hara}

\affil[1]{Department of Mathematical Sciences, Norwegian University of Science and Technology, Norway}
\affil[2]{Center for Biodiversity Dynamics, Norwegian University of Science and Technology, Norway}
\affil[3]{UK Centre for Ecology \& Hydrology, Wallingford, UK.}

\date{}
\begin{document}
\maketitle
\begin{abstract}
\noindent 1. Temporal trends in species distributions are necessary for monitoring changes in biodiversity, which aids policymakers and conservationists in making informed decisions. Dynamic species distribution models are often fitted to ecological time series data using Markov Chain Monte  Carlo algorithms to produce these temporal trends. However, the fitted models can be time-consuming to produce and run, making it inefficient to refit them as new observations become available.
\newline 2. We propose an algorithm that updates model parameters and the latent state distribution (e.g. true occupancy) using the saved information from a previously fitted model. This algorithm capitalises on the strength of importance sampling to generate new posterior samples of interest by updating the model output. The algorithm was validated with simulation studies on linear Gaussian state space models and occupancy models, and we applied the framework to Crested Tits in Switzerland and Yellow Meadow Ants in the UK.
\newline 3. We found that models updated with the proposed algorithm captured the true model parameters and latent state values as good as the models refitted to the expanded dataset. Moreover, the updated models were much faster to run and preserved the trajectory of the derived quantities.
\newline 4. The proposed approach serves as an alternative to conventional methods for updating state-space models (SSMs), and it is most beneficial when the fitted SSMs have a long run time. Overall, we provide a Monte Carlo algorithm to efficiently update complex models, a key issue in developing biodiversity models and indicators.
\newline \textbf{keywords}: biodiversity
monitoring surveys, dynamic species distribution models, importance sampling, particle MCMC, state space models.
\end{abstract}
\ifdefined\Shaded\renewenvironment{Shaded}{\begin{tcolorbox}[interior hidden, enhanced, frame hidden, borderline west={3pt}{0pt}{shadecolor}, breakable, sharp corners, boxrule=0pt]}{\end{tcolorbox}}\fi

\hypertarget{introduction}{%
\section{Introduction}\label{introduction}}

Biodiversity monitoring programs are essential to obtain data for conservation and management decisions. Monitoring programs such as the North American Breeding Bird Survey \citep{bystrak1981north} and the UK Butterfly Monitoring Scheme (\url{https://ukbms.org/}) have generated ecological time series data that have been used to model animal movement \citep{cleasby2019using}, population dynamics \citep{rivot2004bayesian, kery2020applied}, and a range of other ecological properties. Data from mass participation schemes is often used to assess changes in the distributions of species over time \citep{outhwaite2020complex}.  

Ecologists use a wide range of modelling techniques to make inferences on the spatial and temporal patterns in biodiversity from monitoring data. Such models are often developed as state-space models (SSMs) to account for the observation and state processes and to potentially include temporal autocorrelation \citep{doucet2001introduction, kery2020applied, auger2020introduction}. The state or ecological process model captures the dynamics of the state variable (e.g. occupancy or abundance, and is usually unknown and treated as a latent variable) over time, whilst the observation model describes the stochastic dependence of the observed data on the state model. Model parameters and the distribution of the latent state are estimated from these SSMs to provide inferences and predictions about the latent state and the drivers of the ecological and observation processes \citep{newman2023state, durbin2012time}. 

SSMs have gained popularity in applied ecology owing to their flexibility and the availability of packages that provide Monte Carlo algorithms (such as Markov Chain Monte Carlo (MCMC) and sequential Monte Carlo or particle filtering algorithms) to fit them \citep{auger2016state}. The MCMC approaches simulate parameters of interest from their posterior distribution whereas the particle filters simulate from a proposal distribution (also known as importance function) and re-weigh the generated samples to approximate the posterior distribution of the state variables \citep{doucet2001introduction,fearnhead2002markov}. The MCMC approaches are frequently used \citep{auger2016state}, in part due to the availability of generic MCMC frameworks such as BUGS \citep{lunn2000winbugs, lunn2009bugs}, JAGS \citep{depaoli2016just}, Stan \citep{carpenter2017stan}, NIMBLE \citep{de2017programming} and GRETA \citep{golding2019greta}. 

On the other hand, particle filters have been less popular, although R packages such as \textit{nimbleSMC} \citep{michaud2021sequential}, \textit{rbiips} \citep{todeschini2014biips} and \textit{rbi} \citep{murray2013bayesian} (see \citealp{newman2023state} for a review of packages for various particle filter algorithms) have improved the accessibility of some particle filter algorithms to fit SSMs in applied ecology. These particle filter algorithms include bootstrap particle filters \citep{gordon1993novel}, auxiliary particle filters \citep{pitt1999filtering}, Lui-West particle filters \citep{liu2001combined}, ensemble Kalman filter \citep{evensen2003ensemble}, and particle learning \citep{carvalho2010particle}. Further advances in fitting these SSMs have led to the proposal of particle MCMC \citep[pMCMC; ][]{andrieu2010particle}, which allows the joint sampling from the posterior distribution of the latent states and model parameters \citep{andrieu2010particle, gao2012sequential, michaud2021sequential}. The pMCMC algorithm combines particle filter algorithms and the MCMC approach by estimating the posterior distribution of the latent state with particle filter algorithms and that of the model parameters with MCMC \citep{newman2023state, andrieu2010particle}.

Although SSMs have gained widespread use in the quantitative ecology literature, they remain under-used in biodiversity monitoring programs. One reason is that it is computationally expensive (longer run times and use of computational resources) to fit these SSMs using Monte Carlo algorithms \citep{dobson2023dynamicsdm, kery2013analysing}. This can be due to the multimodal joint likelihood surfaces and negligible posterior correlations at high-dimensional parameter space of some SSMs \citep[especially high-dimensional models and non-linear and non-Gaussian SSMs \citep{wang2007latent, peters2010ecological, auger2020introduction} such as the population demographic and dynamic occupancy models we describe in sections \ref{simulation-studies} and \ref{case-study}]{polansky2009likelihood,peters2010ecological}. Consequently, the algorithms used to fit these SSMs can become stuck at certain modes of the likelihood function \citep{andrieu2010particle, peters2010ecological}, which leads to poor or slow convergence. This often makes SSMs time-consuming to fit or wholly impractical. Another reason for the rare use of Monte Carlo algorithms to fit SSMs is the problems for model validation and assessment posed by complex SSM structures \citep{auger2021guide}. The problem of interest to us in this study is the computational expensiveness in fitting the SSMs. This problem of computational time is exacerbated as new biodiversity data regularly becomes available as new records are made, and models need to be refitted with an updated dataset to reflect potentially rapid changes in species distributions \citep{arulampalam2002tutorial}. 

In an attempt to reduce the computational expensiveness of re-fitting SSMs, alternative methods are used to fit sequentially generated time series data. One such alternative approach is fitting independent models to each time slice \citep{kery2020applied, cervantes2023birdie}. Although it may be fast to fit the independent SDMs, such an approach fails to capture the temporal autocorrelation in the data and can lead to biased inferences on state variables of interest such as abundance and occupancy \citep{kery2020applied}. The biased inference of state variables occurs because the independent model requires an intercept for each year and thus produces a highly over-parameterised model \citep{kery2020applied}. Other models account for the temporal variation in the ecological process by specifying the temporal effect as a random effect with an autocovariance structure \citep[for example, year-stratified N-Mixture model; ][]{kery2020applied}; or alternatively by using conventional time series models like the autoregressive integrated moving average model \citep{kong2021spatial}.

This study proposes a computationally efficient algorithm to re-fit SSMs to an updated dataset by sequentially updating an already-fitted SSM with new observations. The proposed algorithm is a modified pMCMC algorithm that keeps the posterior distribution of the latent states and model parameters from the already-fitted model, but approximates the latent state distribution at the new time with a particle filter algorithm and updates the model parameter posterior distribution via the MCMC algorithm. To achieve this, we modify the particle filtering algorithms (specifically the bootstrap and auxiliary particle filters) and MCMC samplers from the R package \textit{nimbleSMC} \citep{michaud2021sequential} to a) store the necessary information from the previous model (fitted to observations up to time $t$) that will be needed to update the parameters when new data is obtained at time $T = t+n$ (where $n$ is the number of years in the future data is obtained); b) estimate and update the posterior distributions of the latent states and model parameters for the new time steps using the stored information in a). We expect the proposed algorithm to reduce the computational time of re-fitting SSMs with data obtained in the future whilst providing a reasonable estimate of the model parameters and latent state at the future time. We assess the performance of the proposed algorithm compared to the conventional SMC and MCMC methods through simulation studies with a linear Gaussian SSM and dynamic occupancy model \citep{kery2020applied}; as well as real-world examples on Yellow Meadow Ant (\textit{Lasius flavus}) and Crested Tits from the Swiss Breeding Bird Survey \citep[accessed from ][]{ahmbook} by comparing their effective sample sizes, computational time and Monte Carlo standard errors.

\section{Materials and Methods}\label{materials-and-methods}

\subsection{Intuition behind the proposed algorithm}\label{Framework}
Our starting point is to assume that we have an existing SSM (fitted using MCMC) and a set of more recent observations than those in the existing model. We aim to combine these statistically to update the existing model. We refer to the existing model we intend to update as the \textbf{reduced model} (here, the term "reduced" is used to indicate that the existing model that is updated will have more observations and the time steps will be more than the existing model). We also refer to updating the existing model with additional observations using the proposed algorithm as the \textbf{updating process} and the existing model updated using the proposed algorithm as the \textbf{updated model}. Therefore, both reduced and updated models are the same SSM but fitted with different observations and/or Monte Carlo methods.

We can assume that with our reduced model (up to time $t$), we have estimated the distribution of latent states, $x_{1:t}$ and other parameters, $\theta$, using MCMC. This means that we draw them from their posterior distribution, $p(x_{1:t}, \theta | y_{1:t})$, so each draw is a vector $\boldsymbol{\psi}^{(i)} = (x^{(i)}_{1:t}, \theta^{(i)} | y_{1:t}), i=1,..., N$, where $N$ is the number of MCMC samples. These draws are sometimes called particles. If posterior summaries are of interest, say the mean of $x_4$, it can be calculated as $\frac{1}{N}\sum_{i=1}^N x_4^{(i)}$.

An alternative to MCMC would be to sample uniformly from $\boldsymbol{\psi}$, $M$ times. Then for each draw, a weight $w^{(i)} \propto p( y_{1:t}| {x}_{1:t}^{(i)}, \theta^{(i)} )$, $i = 1, 2, \ldots, M$, is assigned to it. Now, if we want to estimate the mean of $x_4$ as we did above, we calculate $\frac{1}{W}\sum_{i=1}^M w^{(i)} x_4^{(i)}$, where $W = \sum_{i=1}^M w^{(i)}$. This is the underlying idea behind SMC methods. Without going into the details, we can draw $\boldsymbol{\psi}$ from any reasonable distribution and then find the correct way to calculate the weights, $w$ (refer to \citealt{newman2023state,auger2021guide} for further details).

The idea in this study is that we draw $\boldsymbol{\psi}$ from the posterior distribution for our reduced model, and calculate its associated importance weight. Then if we add more data, $y_{t+1}$, we could simply calculate a new weight iteratively as, $w_{t+1} \propto w_{t} \times p(y_{t+1}|\boldsymbol{\psi})$. This would be a simple bootstrap particle filter algorithm. Whilst it will often work, if too much data is added, most particles will have a very small weight. This makes the distributions' calculations more error-prone. Thus, we need some way of regenerating particles, so we still have many with a substantial weight. Amongst the alternative particle filters we could employ in such an instance, we look at the auxiliary particle filter- Further details are provided in sections \ref{model-set-up} and \ref{updating-process-using-pmcmc}.

\subsection{Set-up of proposed Monte Carlo algorithm}\label{model-set-up}

For $T$ time points, let $y_{1:T} = \{y_1, y_2, \ldots, y_T\}$, where \(y_t\) is a
(\(k \times 1\)) vector, denote observations from a (multivariate) observed time series dataset, e.g. counts of a species at one or more locations. These observations depend on latent states \(x_{1:T} = \{x_1, x_2, \ldots, x_T\}\), which represent the unobserved ecological process, e.g. the actual abundance of the species. These latent states are assumed to have a first-order Markov structure (the latent state at time \(t\) depends on the latent state at time \(t-1\) only). The observations at each time \(t\), \(y_t\), given the latent state at that time \(x_t\), are independent of previous observations and states.

In summary, we have the following information for the SSM framework:
\begin{equation}\label{SSM}
\begin{split}
\text{Intial state distribution} :& \quad p(x_1| \theta); \quad t = 1 \\
\text{State model} :& \quad p(x_t| x_{t-1},\theta); \quad t = 2, 3, \ldots, T \\
\text{Observation model} :& \quad p(y_t| x_{t},\theta); \quad t = 2, 3, \ldots, T \\
\text{Prior distribution of parameters} :& \quad p(\theta),
\end{split}
\end{equation}
where \(\theta\) are model parameters (assumed to be stochastic in the SSM defined in equation \ref{SSM}). It should be noted that the parameters \(\theta\) could be reformulated as year-dependent (for example, an autoregressive year effect) or constant over time.

\begin{table}[htbp!]
\centering
\begin{tabular}{ | m{5em} | m{35em}| } 
 \hline
 Notation & Full definition  \\ [0.5ex] 
 \hline\hline 
  SSM & State space model \\[1ex] 
  MCMC & Markov Chain Monte Carlo (a Monte Carlo approach to fit state space model by drawing samples from the posterior distribution of the parameters) \\[2ex] 
  SMC & Sequential Monte Carlo (another Monte Carlo approach to fit state space models by drawing samples from the prior distribution and re-sampling them using their importance weights)\\[2ex] 
  pMCMC & Particle MCMC (a Monte Carlo approach that utilizes the MCMC approach to sample the model parameters and SMC to approximate data likelihood)\\[2ex] 
   RM & Reduced model (The state-space model that has already been fitted to the observed data to time $t$ using MCMC)\\[2ex] 
UM & Updated model (SSM fitted with the proposed Monte Carlo algorithm to observed data from time $1$ to $T$ using the stored posterior samples from the reduced model)\\[2ex] 
BM & Baseline model (SSM fitted to the observed data from time $1$ to $T$ using MCMC. The results from the updated models are compared to those from the baseline models)\\[2ex] 
 $t$ & Time point used to fit the reduced model\\[1ex] 
 $y_{1:T}$ & Observed data from time $t = 1$ to $t = T$ \\[1ex]
 $x_{1:T}$ & Latent states distribution from time $t = 1$ to $t = T$\\ [1ex]
 $\theta$ & Model parameters \\ [1ex]
$N$ & Number of samples from the posterior distribution of the reduced model (after thinning the left-over samples from the burn-in phase) \\[1ex]
$M$ & Number of sequential Monte Carlo particles used (Not necessarily equal to N) \\[1ex]
 \hline

\end{tabular}
 \caption{Notations used in this manuscript and its full definition.}
\label{table:modelabbrev}
\end{table}

\begin{figure}[htbp!]
\begin{tikzpicture}
    \draw[black,thick,latex-latex] (0,0) -- (15,0)
    node[pos=0,label=above:\textcolor{red}{$Time: 1$}]{}
    node[pos=0.25,text=blue,label=below:\textcolor{black}{$ \underbrace{\begin{bmatrix}
{x}_{1}^{(1)} & {x}_{2}^{(1)} & \ldots &{x}_{t}^{(1)} \\
{x}_{1}^{(2)} & {x}_{2}^{(2)} & \ldots & {x}_{t}^{(2)} \\
\vdots & \vdots & \ldots & \vdots \\
{x}_{1}^{(N)} & {x}_{2}^{(N)} & \ldots & {x}_{t}^{(N)}
\end{bmatrix} 
\begin{bmatrix}
{\theta}_{r}^{(1)} \\
{\theta}_{r}^{(2)}\\
\vdots \\
{\theta}_{r}^{(N)}
\end{bmatrix}}_{\text{MCMC used for State Space models ($\boldsymbol{\psi}_{1:t}$)} }$}]{}
node[pos=0.80,text=blue,label=below:\textcolor{blue}{
$ \underbrace{\begin{bmatrix}
x_{t +1 }^{(1)} &\ldots &x_{T}^{(2)} \\
x_{t +1}^{(2)} & \ldots & x_{T}^{(2)} \\
\vdots &  \ldots & \vdots \\
x_{t +1}^{(N)} & \ldots & x_{T}^{(N)}
\end{bmatrix}
\begin{bmatrix}
\theta_{upd}^{(1)} \\
\theta_{upd}^{(2)}\\
\vdots \\
\theta_{upd}^{(N)}
\end{bmatrix}}_{\text{SMC used to update latent states and parameters}}$}]{}
    node[pos=0.75,text=blue,label=above:\textcolor{blue}{New observation: $y_{t +1}, \ldots, y_{T}$}]{}
        node[pos=0.25,text=black,label=above:\textcolor{black}{Old observation: $y_{1}, \ldots, y_{t}$}]{}
    node[pos=0.50,fill=black,text=black,label=above:\textcolor{red}{$t$}]{}
    node[pos=1,fill=blue,text=blue,label=above:\textcolor{red}{$T$}]{};
  \end{tikzpicture}
  \caption{\label{framework} Framework of saved posterior samples of latent states and model parameters from the reduced model and the posterior samples to be obtained from the updated model. In this framework, we provide a notation difference between the estimates of $\theta$ from the reduced and updated models. $\theta_r^{(j)}$ and $x^{(j)}_{1:t}$ (for $j = 1, 2, \ldots, N$) are draws from the posterior distribution of $\theta$ and $x_{1:t}$ respectively from the already-fitted SSM to the observations $y_{1:t} = \{y_1, y_2, \ldots, y_{t} \}$. $\theta_{upd}^{(j)}$ and $x^{(j)}_{t+1:T}$ (for $j = 1, 2, \ldots, N$) are posterior samples of parameters and latent states respectively we are interested in drawing from the posterior distribution $p(x_{1:T}, \theta |y_{1:T})$ after the new set of observations $y_{{t +1}: T} = \{y_{t +1}, y_{t +2}, \ldots, y_{T} \} $ have been obtained. $N$ is the number of samples drawn from the posterior distribution.}
  \end{figure}
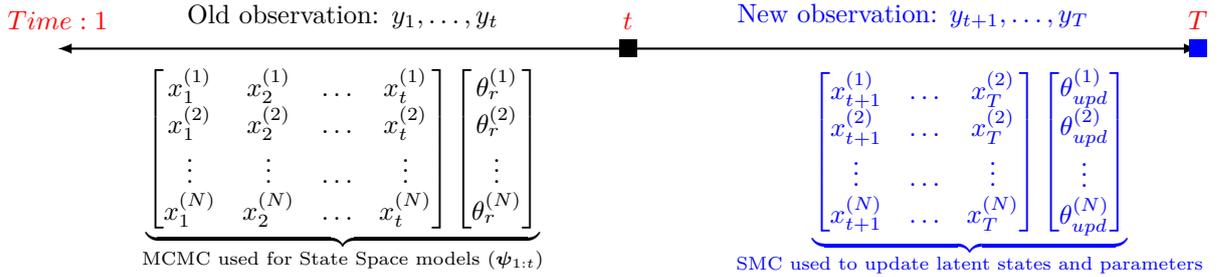

We assume the SSM has already been fitted to the data, excluding the current time point (say at time $t < T$) using MCMC. Consequently, stored posterior samples of latent states and model parameters are available. Let $\boldsymbol{\psi}_{1:t}^{(i)} = (x^{(i)}_{1:t}, \theta^{(i)} | y_{1:t})$, $i=1,..., N$,  be the draw from the posterior distribution of the reduced model (i.e. stored MCMC samples). We are interested in approximating $p(x_{1:T}, \theta |y_{1:T})$ by drawing new samples for time $t+1$ to $T$ whilst leveraging on the posterior distribution we have for time up to $t$ (i.e. $\boldsymbol{\psi}_{1:t}$). The structure of the stored and the expected posterior samples of the latent states and model parameters are presented in Figure \ref{framework}.

\subsection{Monte Carlo algorithm for the updating process}\label{updating-process-using-pmcmc}

Our framework relies on some form of particle filter (in this study, we use the pMCMC proposed by \citeauthor{andrieu2010particle} because we approximate the joint posterior distribution of latent states and other parameters) for the updating process. pMCMC uses a Metropolis-Hastings step in generating new values of model parameters whilst approximating the model likelihood with particle algorithms (see \citet{andrieu2010particle} for details on pMCMC and its implementation). Therefore, the following steps are used to update the existing SSM:

\begin{itemize}
\item for each MCMC iteration $j = 1, 2, \ldots, N$
\item \textbf{Step 1:} for time steps $1$ to $t$, copy the same values of $\boldsymbol{\psi}_{1:t}^{(j)}$ into the Monte Carlo algorithm $M$ times. Since these values are duplicated as $M$ particles, we calculate the importance weight once and duplicate them $M$ times.
\item \textbf{Step 2:} for time steps $t+1$ to $T$, simulate $M$ particles for the latent state $x_{t+1:
T}$ from the particle filter algorithm (for example, bootstrap particle filter), estimate their associated importance weights and approximate the model likelihood. Note that the approximation of the model likelihood uses the estimated weights from time steps $1$ to $T$.
\item \textbf{Step 3:} Using this estimated model likelihood, we simulate values of the model parameters $\theta$ from their posterior distribution using an MCMC algorithm. 
\end{itemize}
We provide details of each step described above in sections \ref{updating-the-sequential-monte-carlo-algorithm} and \ref{updating-the-metroplis-hasting-approach}.

\subsubsection{Updating the particle filter
algorithm}\label{updating-the-sequential-monte-carlo-algorithm}

Each draw of the latent states for time steps $1, 2, \ldots, t$ from the reduced model are replicated $M$ times into the particle filter (and the importance weight is also calculated once and replicated $M$ times). For time steps $ s= t+1, t+2, \ldots, T$, we simulate $x_{s}$ from its prior distribution and re-sample the $M$ particles using the importance weights $w_t$. In summary, for the $j^{th}$ ($j = 1, 2, \ldots, N$ samples) draw of $x_{1:t}$ and $\theta$ from the posterior distribution of the reduced model (which we denote as $\hat{x}_{1:t}^{(j)}$ and $\hat{\theta}_{r}^{(j)}$ respectively), the following information is used for the particle algorithms in the updating process: 
\begin{equation}\label{weights} 
\left\{
\begin{array}{lll}
      w_s^{(i)} = w_{s-1}^{(i)} \times p(y_{1:s}|\hat{x}_{1:s}^{(j)}, \hat{\theta}_{r}^{(i)}) & x_{s}^{(i)} = \hat{x}_{s}^{(j)} & \text{for } s = 1, 2, \ldots, t \text{ and } 1\leq  i \leq M\\
      w_s^{(i)} \text{ defined by any SMC algorithm }  & x_{s}^{(i)} \text{ from SMC}& \text{for } s=t+1, t+2, \ldots, T \text{ and } 1\leq  i \leq M\\
\end{array} 
\right. 
\end{equation} 
where $w_s^{(i)}$ is the importance weight for the $i^{th}$ particle and $x_{s}^{(i)}$ is the latent state distribution for $i = 1, 2, \ldots, M$ particles. 

This study focuses on two particle filter algorithms: bootstrap and auxiliary particle filter algorithms. We briefly introduce both particle filter algorithms in Supplementary Information 1. We also implement our proposed algorithm by copying and editing the bootstrap and auxiliary particle filter algorithm implemented in the R-package \textit{nimbleSMC} \citep{michaud2021sequentialPackage}. The bootstrap and auxiliary particle algorithms we implement are presented in Supplementary Information One, Algorithms 1 and 2, respectively. 

\subsubsection{Updating the Metropolis-Hasting
algorithm}\label{updating-the-metroplis-hasting-approach}

We update the model parameters in the SSM ($\theta$) by proposing values from a distribution that depends on its posterior distribution from the reduced model. That is, the proposal distribution used is defined as $\pi(\theta^{\star}| {\theta}_r)$, where $\theta^{\star}$ is the candidate or proposed model parameter value and ${\theta}_r$ is a draw of $\theta$ from its posterior distribution for reduced model. This proposed value $\theta^{\star}$ is accepted or rejected using an acceptance ratio that depends on the likelihood of the model. This model likelihood, however, is approximated from the particle filtering algorithm in section \ref{updating-the-sequential-monte-carlo-algorithm}. The approximation depends on the importance weights from the particle filter algorithms defined in equation \eqref{weights}.

Using this proposal distribution and importance weights, we describe a Metropolis-Hastings acceptance ratio for the updating process (\(MHAR_{upd}\)) as follows:
\begin{equation}\label{MHARupd}
MHAR_{upd} = \frac{p(\theta^{\star})\pi({\theta}_{r}|\theta^{\star})\prod_{s=t +1}^{T}\sum_{i = 1}^{M} w_{s|\theta^{\star}}^{(i)}}{p({\theta}_{r})\pi(\theta^{\star}|{\theta}_{r})\prod_{s=t+1}^{T}\sum_{i = 1}^{M} w_{s|{\theta}_{r}}^{(i)}};
\end{equation} where \(w_{s|\theta^{\star}}^{(i)}\) is the \(i^{th}\) importance weight at time \(s\) given the value of \(\theta^{\star}\), \(w_{s|{\theta}_{r}}^{(i)}\) is the \(i^{th}\) importance weight at time \(s\) given the value of \({\theta}_{r}\), \(p(\theta^{\star})\) and \(p({\theta}_{r})\) are the prior
distribution of \(\theta^{\star}\) and \({\theta}_{r}\) respectively and \(\pi(.|.)\) is the proposal distribution for the parameters \(\theta\). The proposed value is accepted with probability min(1, \(MHAR_{upd}\)) and $\theta_{upd}$ is set to $\theta^{\star}$, otherwise  $\theta_{upd} $ is set to ${\theta}_{r}$. See Supplementary Information One for details on how the updated Metropolis-Hastings acceptance ratio defined in equation \eqref{MHARupd} was derived and Supplementary Information One, Algorithm 3 for details on changes in the Metropolis-Hasting algorithm. 

\textit{nimbleSMC} jointly samples all model parameters, including any hyperparameters, by proposing $\theta$ from a normal distribution (i.e. random walk block sampling algorithm). This can reduce the efficiency of the particle MCMC, especially when the model parameters are over-parameterised \citep{tibbits2014automated}. To improve the efficiency of the updating algorithms, we sample the hyperparameters separately from the other model parameters. We split the model parameters $\theta = \{\theta_1, \theta_2 \}$ such that the distribution of $\theta_1$ depends on $\theta_2$ (i.e. $p(\theta) = p(\theta_1|\theta_2) p(\theta_2)$). We first draw samples for $\theta_2$ by proposing candidate values from a normal distribution centred around its MCMC draw from the posterior distribution of the reduced model (i.e. we choose $\pi(\theta_2^{\star}| \theta_{r_2}) = N_p(\theta_{r_2}, \delta_2 \Sigma_2)$ where $\theta_{r_2}$ is a draw from the posterior distribution of $\theta_2$ from the reduced model, $\delta_2$ is the scale tuning parameter, $\Sigma_2$ is a ($p \times p$) variance-covariance matrix for the simulation and $p$ is the number of parameters in the set of model parameters $\theta_2$). Given the draw of $\theta_2$, we sample $\theta_1$ by proposing candidate values from normal distribution centred around its MCMC draw from the posterior distribution from the reduced model and the sample of $\theta_2$ (i.e. we choose $\pi(\theta_1^{\star}| \theta_{r_1}, \theta_2) = N_p(\theta_{1}|\theta_2, \delta_1 \Sigma_1)$ where $\theta_{r_1}$ is a draw from the posterior distribution of $\theta_1$ from the reduced model, $\delta_2$ is the scale tuning parameter, $\Sigma_1$ is a ($p \times p$) variance-covariance matrix for the simulation of $\theta_1^{\star}$ and $p$ is the number of parameters in the set of model parameters $\theta_1$). In this study, we choose $\delta_1 = \delta_2 < 0.5$, $\Sigma_1$ and $\Sigma_2$ to be identity matrices unless otherwise stated. We provide details of this sampling process in Supplementary Information One, Algorithm 4. 

By the definition of the proposed Monte Carlo algorithm, the draws of latent state $x_{t+1:T}$ for each iteration $j \in{1, 2, \ldots, N}$ is a forward simulation when the candidate values of $\theta$ are not accepted (i.e. $x_{t+1:T}$ is simply a simulation from its prior distribution given the parameter values from the reduced model). When the candidate values ($\theta^{\star}$) are accepted, then $x_{t+1:T}$ is simulated from the prior distribution given $\theta^{\star}$.
\subsection{Simulation Studies}\label{simulation-studies}

To validate our proposed Monte Carlo algorithm and evaluate some of its properties, we perform two simulation studies. We are specifically interested in how well we estimate the posterior distribution of $x_{t+1:T}$ and $\theta$ from the updating process, the effect of the number of time steps $t$ used to fit the reduced model on the updating process (we fix $T$ and vary $t$), and the performance of the Metropolis-Hastings algorithm for the updating process.

 In the first study, we simulate a time series with $90$ data points from a linear Gaussian SSM and use our proposed algorithm to fit the model to the simulated data (section \ref{simulation-study-one}). We use the median bias and Monte Carlo standard error (MCSE) as our model performance metrics. In the second simulation, we choose a more realistic ecological SSM to simulate data from, specifically the dynamic occupancy model (section \ref{simulation-study-two}). We examine the trends of functions of the latent state, the MCMC convergence using the traceplot and the performance of the proposed samplers by estimating the effective sample size and efficiency  (i.e., the number of independent samples per unit time) from the chains of the updated models. The trace plots and effective sample samples are obtained using the R-package \textit{ggmcmc} \citep{ggmcmc}.

In both simulations, we fit a full model using MCMC (and study one, with pMCMC also using both bootstrap and auxiliary particle filter) to the observed data ($y_{1:T}$). This full model is the baseline we compared to the SSMs fitted with the proposed MC algorithm. We fit a reduced model using the MCMC approach to the observed data up to time $t$, and the reduced model is then updated with data for up to time $T$ using the proposed algorithm. We use the R-packages \textit{nimble} \citep{de2017programming} and \textit{nimbleSMC} \citep{michaud2021sequentialPackage} to fit the models using the MCMC and pMCMC algorithms respectively. A summary of the models fitted, the type of model, and the Monte Carlo method used are summarised in Table \ref{table:modelSummary}.

\begin{table}[h!]
\centering
\begin{tabular}{||p{1cm} | p{3cm} | p{4cm}| p{5cm}||} 
 \hline
 Model name & Full definition & Monte Carlo method & Model type \\ [0.5ex] 
 \hline\hline
%ABSC & Auxiliary baseline SMC & particle MCMC with Auxiliary PF for SMC &  \multirow{3}{4em}{Baseline model} \\ [1ex]
 % BBSC & Bootstrap baseline SMC & particle MCMC with Bootstrap PF for SMC &  \\ [1ex]
BMC &  Baseline MCMC & MCMC & Full model fitted to entire observations up to time $T$ and used as a baseline to compare our proposed algorithm. \\ [1ex]
 \hline 
  %ARSC & Auxiliary reduced SMC & Auxiliary PF &  \multirow{4}{4em}{Reduced model}\\[1ex] 
 %BRSC & Bootstrap reduced SMC & Bootstrap PF &  \\[1ex] 
 RMC  & Reduced MCMC & MCMC & SSM fitted to observations up to time $t$ \\[1ex] 
 \hline 
  %AUSC  & Auxiliary updated SMC &  Auxiliary PF & \multirow{4}{4em}{Updated model} \\ [1ex] 
  %BUSC  & Bootstrap updated SMC &  Bootstrap PF &  \\ [1ex] 
 AUMC & Auxiliary Updated MCMC & particle MCMC with Auxiliary PF & \multirow{2}{5cm}{Updated reduced model fitted to entire observations up to time $T$ using our proposed algorithm}\\[2ex] 
 BUMC & Bootstrap Updated MCMC & particle MCMC with Bootstrap PF &  \\[1ex] 

 \hline
\end{tabular}
\caption{Model names and types with the Monte Carlo method used for the examples. The model names are abbreviations of the model type and the Monte Carlo method used. For example, "AUMC" takes the first letter of the Monte Carlo method used (auxiliary particle filter), the second letter in the abbreviated name is taken from the model type (updated model), and the last two letters indicate that the model was fitted using the Monte Carlo approach. }
\label{table:modelSummary}
\end{table}

\subsubsection{Linear Gaussian state-space
model}\label{simulation-study-one}

We use a modified form of the linear Gaussian SSM in the NIMBLE manual \citep{nimblepackage} for our first illustration. We have changed the distribution of the initial latent state and the observation process model. The linear Gaussian SSM is chosen because it is simpler to fit and has a relatively short run time, making it ideal for exploring the properties of the proposed Monte Carlo algorithm. 

The linear Gaussian SSM we aim to fit is defined as: \begin{equation}\label{LGSSM}
\begin{split}
%x_1 &\sim N \big(\frac{b}{1-a}, 1 \big)\quad \text{for} \quad t = 1,\\
x_1 &\sim N \big(a, 1 \big)\quad \text{for} \quad t = 1,\\
y_1 &\sim N(x_1, 1) \quad \text{for} \quad t = 1,\\
x_t &\sim N(ax_{t-1}, 1) \quad \text{for} \quad 2 \leq t \leq 50, \\
y_t &\sim N(c x_t, 1) \quad \text{for} \quad 2 \leq t \leq 50,\\
\end{split}.
\end{equation} where \(x_{1:50}\) is the latent state variable,
\(y_{1:50}\) is the observed data and \(a\) and \(c\) are the model parameters to be estimated. 

We simulate $90$ replicate datasets from this model using the following parameter values: \(a = 0.5\) and \(c = 1\). To assess the effect of the estimation of the posterior distribution from the reduced model on the updated model, the number of time steps (\(t\)) used to fit the reduced model is varied. Specifically, we choose \(t =5, 10, 20, 49\). For each \(t\), the models summarised in Table \ref{table:modelSummary} are fitted. For all models, we run three chains with  \(30000\) iterations with the initial \(20000\) used as burn-in samples and keep the remaining samples. The same number of particles, $M = 1000$, is used to fit all the models in this simulation study. For each simulated dataset, we estimate the following metrics to assess the performance of the algorithms:

\begin{equation}
    \begin{split}
        Bias(\theta) &= \hat{\theta} - \theta\\
        Bias(x_{1:50})&= \frac{1}{50}\sum_{i=1}^{50}(\hat{x}_i - x_i),
    \end{split}
\end{equation}
where $\hat{\theta}$ and $\hat{x}$ are the posterior mean estimated from the fitted models. We summarise each of these metrics over the replicated datasets by presenting the median of these estimates.

\subsubsection{Dynamic occupancy models}\label{simulation-study-two}

The second example we use is a dynamic occupancy model described in Chapters 4.2 - 4.4 in \cite{kery2020applied}. This model is used to explore the computational time of the fitted models, the efficiency of the proposed Metropolis-Hastings algorithm (which is used to sample the model parameters) and how well the updating process captures the ecological process trend over time. 

Dynamic occupancy models capture changes in species presence/absence and are the most natural choice to model (meta)population dynamics where individuals cannot be identified individually \citep{mackenzie2003estimating, mackenzie2012investigating}. They allow us to model spatio-temporal population changes using parameters (which can be functions of population change drivers) that explicitly describe the underlying process \citep{kery2020applied}. In this example, it is of interest to model the spatio-temporal variation in the presence/absence state (\(\textbf{z}\)) at a site, where \(\textbf{z}\) is a site by time matrix of state values. It is also interesting to model the variation in the persistence and detection probabilities, which we assessed by modelling covariate effects. 

The model we aim to fit (similar to what is described in pages $212 -213$ of \cite{kery2020applied}) is defined in equation \eqref{dynamicOccupancy}:
\begin{equation} \label{dynamicOccupancy}
\begin{split}
\text{Initial state: }z_{i,1} &\sim Bernoulli (\psi_i)\\
\text{State dynamics: } z_{i,t} &\sim Bernoulli (z_{i, t-1}  \phi_{i, t-1} + (1 - z_{i, t-1}) \gamma) \\
\text{Observation process: }   y_{i,j,t}| z_{i,t} &\sim Bernoulli (z_{i, t} \times p_{i, j, t})\\
logit(p_{i,j,t}) &= \alpha^{p} + \beta^p* windSpeed_{i,j,t}\\
logit(\psi_{i}) &= \alpha^{\psi} + \beta^{\psi}*elevation_{i}\\
logit(\phi_{i,t}) &= \alpha^{\phi} + \beta^{\phi}*springPrecipitation_{i,t}\\
% logit(\gamma_{i,t}) &= \alpha^{\gamma}_{t} + \beta^{\gamma}_{t}*sizeOfBeak_{i,t}\\
 \text{Prior distributions of model parameters:}\\
\alpha^{p} \sim Normal(0, \sigma^2_{\alpha^{p}}); \quad \alpha^{\psi} \sim Normal(0, 10) ; & \quad \alpha^{\phi} \sim Normal(3, \sigma^2_{\alpha^{\phi}});\\
\beta^{p} \sim Normal(0, \sigma^2_{\beta^{p}}); \quad \beta^{\psi} \sim Normal(0, 10) ; & \quad \beta^{\phi} \sim Normal(1.5, \sigma^2_{\beta^{\phi}});\\
 \gamma \sim Uniform(0.001, 1) \\
 \text{Prior distributions of model hyperparameters:}\\
 \sigma^2_{\alpha^{p}} \sim \textit{truncated Normal} (0, 10, 0.001, \infty) ; & \quad \sigma^2_{\beta^{p}} \sim \textit{truncated Normal} (0, 10, 0.001, \infty) \\
 \sigma^2_{\alpha^{\phi}} \sim \textit{truncated Normal} (0, 10, 0.001, \infty) ; & \quad  \sigma^2_{\beta^{\phi}} \sim \textit{truncated Normal} (0, 10, 0.001, \infty)
\end{split}
\end{equation} 
where \(\psi_{i}\) is the initial occupancy probability at site \(i\), \(\phi_{i,t}\) is the persistence probability (probability that an occupied site at time \(t-1\) will be occupied at
time \(t\)) at site \(i\) and time \(t\), \(\gamma\) is the colonisation probability (probability that an unoccupied site at time \(t-1\) will be occupied at time \(t\)) at site \(i\) and time \(t\), \(p_{i, j, k}\) is the detection probability at site \(i\) during visit \(j\) in year \(t\).

We simulate the occupancy data with \(300\) sites, \(3\) visits and \(T = 30\) years. We also simulate the covariates \emph{windSpeed}, \emph{elevation}, and \emph{springPrecipitation} from a
standard normal distribution. We set the hyperparameters as: \(\sigma_{\alpha^{p}} = 2\), \(\sigma_{\alpha^{\psi}} = 3\), \(\alpha^{\phi} = 2\),  \(\sigma_{\beta^{p}} = 3\), \(\sigma_{\beta^{\psi}} = 2\), \(\sigma_{\beta^{\phi}} = 2\), and we simulate the covariate effects from the distributions defined in equation \eqref{dynamicOccupancy}. Then, we fit the models summarized in Table \ref{table:modelSummary} for $50000$ iterations, use the first $20000$ as burn-in samples, and keep a sixth of the remaining samples. 

Additionally, we vary the number of years ($t$) used to fit the reduced model. We choose $t = 25$ and $t= 29$ years, as we assume that this algorithm would realistically be used to update SSMs for a short period. Also, for models fitted with the proposed Monte Carlo algorithm, we vary the number of particles ($M$) used to fit the models. We anticipate the proposed algorithm would perform better (unbiased estimates of derived quantities of latent states and model parameters) as $M$ increases but with greater computational time. We choose $M=10, 25, 50$ and $100$.

We estimate the realised occupancy probability \citep[$\Psi_{t}^{fs}$; ][]{royle2007bayesian} and monitor its trend over time. The realised occupancy probability is estimated as follows:
\begin{equation}\label{realisedOcc}
    \Psi^{fs}_t = \frac{\sum_{i = 1}^{R} z_{it}}{R},
\end{equation} 
where $t$ is the year index and $R$ is the number of sites. We assess the estimate of the realised occupancy from each of the fitted models by presenting the correlation and bias in realised occupancy (expressed as a percentage), which are estimated as follows:
\begin{equation}\label{Growthrate}
\begin{split}
 \text{Correlation($\Psi$)} &= corr(\hat{\Psi}_{t+1:T}^{fs}, {\Psi}_{t+1:T}^{fs})\\
    \text{Bias($\Psi_T$)} &= \bigg[ \hat{\Psi}^{fs}_{T} -  \Psi^{fs}_{T}  \bigg] \times 100,\\
\end{split}
\end{equation}
where $\hat{\Psi}_{t+1:T}^{fs}$ is the estimated realised occupancy probability from years $t+1$ to the last year ($T$) from the fitted models, ${\Psi}_{t+1:T}^{fs}$ is the actual realised occupancy probability from years $t+1$ to the last year ($T$) from the simulated data set, $\hat{\Psi}^{fs}_{T}$ is the estimated realised occupancy probability for the last year ($T$) from the fitted models and ${\Psi}^{fs}_{T}$ is the realised occupancy probability for the last year ($T$) from the simulated dataset.

\subsection{Case study}\label{case-study}

We also show the use of the proposed algorithm in fitting models to real datasets. The datasets in this study are generated from biodiversity monitoring programs where observations are reported yearly. We assume that derived quantities from these reported observations must be obtained yearly (i.e. we update the existing model just a year ahead).

Specifically, we apply the proposed algorithm to fit SSMs to Crested Tit (\textit{Lophophanes cristatus}) data from Switzerland and Yellow Meadow Ant (\textit{Lasius flavus}) data from the UK. We aim to infer the annual relative abundance of Crested Tits by fitting a demographic SSM with generalized Markovian dynamics and covariates and infer the annual occupancy of Yellow Meadow Ant by fitting a dynamic occupancy model with autoregressive year random effect. Further descriptions of the two models are presented in sections \ref{case-study-one} and \ref{case-study-two}. These two examples aim to further illustrate our models' performance to real datasets, i.e. to show that we have developed an efficient approach to updating monitoring results and species trends. The Yellow Meadow Ant analysis demonstrates how a model fitted using the R package \textit{sparta} \citep{august} can be used, so our approach can be added to existing workflows \citep{boyd2023operational}.

\subsubsection{Demographic SSM with generalized Markovian dynamics and with covariates}\label{case-study-one}

We fit a demographic SSM for Swiss Crested Tits described in Chapter 1.7.2 in \citet{kery2020applied}. The demographic SSM is used to infer the relative abundance of Crested Tits from the Swiss Breeding survey between 1996 and 2016 by explicitly distinguishing between the ecological and observational processes whilst accounting for imperfect detection \citep{kery2020applied}. The data were accessed from the R-package \textit{AHMbook} version 0.2.6 \citep{ahmbook}.

In this SSM, the initial system of the state (abundance) is assumed to be Poisson distributed with expected initial abundance $\lambda$. For the rest of the process, the latent state is assumed to be Poisson distributed with expected abundance governed by the population growth rate $\gamma$. The observed counts are assumed to be Binomial distributed with detection probability $p$ in the observation process. The site-level changes in initial expected abundance and population growth at each site are modelled with elevation and forest cover as covariates, and the site-level detection probability is modelled with date and duration as covariates.

A summary of the SSM with the covariate effects and prior distributions on model (hyper)parameters is described in equation \eqref{example3}:
\begin{equation}\label{example3}
    \begin{split}
\text{Observation Process}: & \quad C_{i,t} \sim Binomial (N_{i,t}, p_{i,t})\\
\text{Initial state distribution}: & \quad N_{i,1} \sim Poisson (\lambda_i)\\
\text{State distribution}: & \quad N_{i,t} \sim Poisson (N_{i,t-1} * \gamma_i + \delta_i)\\
\text{Random immigration process}: & \quad \delta_i \sim Normal (0, \sigma_{\delta}^2)\\
\text{logit}(p_{i,t}) &= \alpha_p + \beta_{p,1}*date_{i,t} + \beta_{p,2}*date_{i,t}^2 + \beta_{p,3}*dur_{i,t} \\
\text{logit}(\lambda_{i,t}) &= \alpha_{\lambda} + \beta_{{\lambda},1}*elev_{i,t} + \beta_{{\lambda},2}*elev_{i,t}^2 + \beta_{{\lambda},3}*forest_{i,t} \\
\text{logit}(\gamma_{i,t}) &= \alpha_{\gamma} + \beta_{{\gamma},1}*elev_{i,t} + \beta_{{\gamma},2}*elev_{i,t}^2 + \beta_{{\gamma},3}*forest_{i,t} \\
 \text{Prior distributions:} & \\
\alpha_p = logit(\mu_p); \quad  \mu_p \sim Uniform(0,1); & \quad  \beta_{p,i} \sim Normal(0,10)\\
\alpha_{\lambda} = logit(\mu_\lambda); \quad  \mu_{\lambda} \sim Uniform(0,1); & \quad  \beta_{\lambda,i} \sim Normal(0,\sigma_{\lambda}^2)\\
\alpha_{\gamma} = logit(\mu_{\gamma}); \quad  \mu_{\gamma} \sim Uniform(0.9,1.1); & \quad  \beta_{\gamma,i} \sim Normal(0,\sigma_{\gamma}^2)\\
\sigma_{\lambda} \sim \text{half-Normal}(0, 10); &\quad \sigma_{\gamma} \sim \text{half-Normal}(0, 10); \\
\sigma_{\rho} \sim \text{half-Normal}(0, 2)
    \end{split}
\end{equation}

We fit the reduced demographic SSM using the observations from the first $17$ years ($1996 - 2015$) out of the $18$ years. We also use the results from this reduced model and all the observations from $1999 - 2016$ to fit the updated demographic SSM. For efficiency, the parameters in the updated models are sampled using the sampler proposed in this study (Supplementary Information One, Algorithm 4). That is, we sample the parameters $\sigma_{\lambda}$, $\sigma_{\gamma}$, $\sigma_{\rho}$,  $\mu_p$ and $\beta_p$ first, and the other model parameters that depend on these parameters are sampled given the sampled values of the former parameters. Using $M = 100$ particles throughout the analysis of the dataset, we run two chains for all models defined in Table \ref{table:modelSummary} with $50000$ iterations, of which $40000$ are used as burn-in samples, and we keep the rest of the samples. We monitor the annual population size and interannual growth rate over the years, with each estimated as: 
\begin{equation}
\begin{split}
 \text{Population size} (N^{\star}_{t}) &= \sum_{i=1}^R N_{it},\\
 \text{Growth rate}(t) &= \frac{N^{\star}_{t+1}}{N^{\star}_{t}},\\
 \end{split}
\end{equation}
where $R$ is the number of sites.

\subsubsection{Occupancy model with \textit{sparta}}\label{case-study-two}

The second case study uses an occupancy model fitted with the R package \textit{sparta} \citep{august}, which is freely available on GitHub  (\url{https://github.com/BiologicalRecordsCentre/sparta}), and it contains various methods to analyse unstructured occurrence records. \textit{sparta} fits models using the MCMC approach via JAGS \cite{depaoli2016just}. Details of the entire workflow for producing periodic estimates of species occupancy using \textit{sparta} are described in \cite{boyd2023operational}.

The UK Centre for Ecology \& Hydrology fits occupancy models for many species using standardized biodiversity data collated from multiple UK or Great Britain-based societies and recording schemes, the iRecord database (\url{https://www.brc.ac.uk/irecord/}) and the Biological Records centre \citep{outhwaite2019annual}. These models take weeks to run across species, with some individual species models taking multiple days \citep{boyd2023operational}, so it is impractical to run the models for all species annually. This case study will fit a model to a single ant species, the Yellow Meadow Ant (\textit{Lasius flavus}), with data collated from $1970$ to $2021$. 

As described in section \ref{simulation-study-two}, occupancy models are used to model species distributions using their presence/absence history. The occupancy model used here is derived from the "random walk" model of \citet{outhwaite2018prior}. This occupancy model accounts for temporal variation in the data with a random walk prior to the year effect and spatial variation with a site random effect. The detection probability was modelled with covariates \textit{datatype2} (categorical specification for a species list of length 2-3) and \textit{datatype3} (categorical specification for a species list of length 4+). See \citet{van2013opportunistic} for further details of the covariate classification and \citet{outhwaite2019annual} for a detailed description of the occupancy model fit here. 

The details of the occupancy model are presented in equation \eqref{example4}:
\begin{equation}\label{example4}
        \begin{split}
\text{Observation Process}: & \quad C_{k} \sim \text{Binomial} (z_{site[k],year[k]}, p_{k})\\
\text{Initial state distribution}: & \quad z_{i,1} \sim \text{Bernoulli} (a_1 + \eta_i)\\
\text{State distribution}: & \quad z_{i,t} \sim \text{Bernoulli}  (a_t + \eta_i)\\
 \text{State Model priors:} & \\
 a_1 \sim \text{Normal}(0, 1000); \quad & a_t \sim \text{Normal}(a_{t-1}, \sigma_{a}^2);  \quad \sigma_{a} \sim \text{half-Cauchy}(0,1)\\ 
 \eta_i \sim \text{Normal}(0, \sigma_{\eta}^2); & \quad \sigma_{\eta} \sim \text{half-Cauchy}(0,1)\\ 
  \text{Linear predictor:} & \\ 
\text{logit}(p_{k}) &= \alpha_p + \beta_{p,1}*datatype2_{i,t} + \beta_{p,2}*datatype3_{i,t} \\
 \text{Prior distributions:} & \\
\alpha_p = \text{Normal}(\mu_p, \sigma_{p}^2); \quad  \mu_p \sim \text{Normal}(0, 100); & \quad  \beta_{p,i} \sim \text{Normal}(0,100)\\
\sigma_{p} \sim \text{half-Cauchy}(0, 1)
    \end{split}
\end{equation}
for $i = 1, 2, \ldots, R$ sites; for $t = 1, 2, \ldots, T$ years; and $i = 1, 2, \ldots, K$ visits. 

The occupancy model is fitted to the observed data up to $2020$ using the \texttt{occDetFunc} function in the R-package \textit{sparta}, and this is the baseline model we compare the other models with. This baseline model took $40$ minutes to run. We also fit the occupancy model to the same observed data using MCMC via the NIMBLE platform, to provide a comparison in the computation times due to the implementation software. This baseline model took over five hours to run.

Similarly, we fit the occupancy model to the observed data up to $2019$ using both the \texttt{occDetFunc} function in the R-package \textit{sparta} (which took $30$ minutes to run) and MCMC via the NIMBLE platform (which took over four hours to run). We fit the model with NIMBLE only to provide a comparison of computational times due to differences in software implementation, as the results from both were comparable. 

 We then update the fitted occupancy model using the R-package \textit{sparta} (reduced model) with the data observed up to $2020$ using the proposed algorithm (this is the updated model).  We sample the model parameters for the updated models from the proposed sampler described in Supplementary Information One, Algorithm 4 for efficiency. We run three chains for all models with $32000$ iterations, of which $30000$ are used as burn-in samples, and we keep every other sample of the remaining samples. Due to the long computational time used to fit the reduced model, we use $M=50$ particles to fit the updated models. We monitor the realised occupancy probability  (defined in equation \ref{realisedOcc}) at $2020$.

\section{Results}
\subsection{Simulation study}
\subsubsection{Estimating posterior distribution}

\begin{figure}[htbp!]
\includegraphics{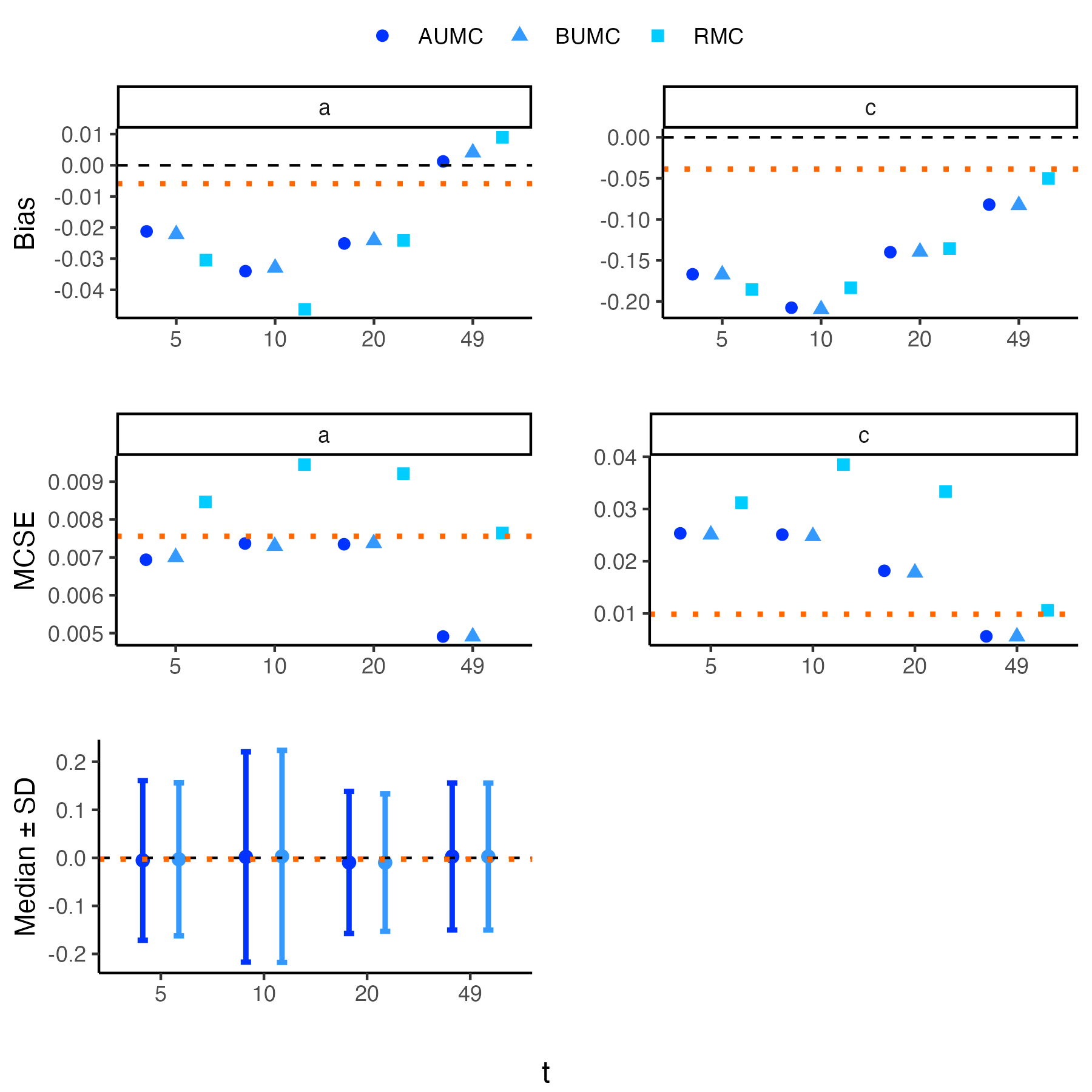}
\caption{\label{lgssm} First row: Median bias in model parameters (a, c); second row: Monte Carlo standard error (MCSE) of the model parameters (a, c); last row: distribution of mean bias of the latent state distribution $x_{1:50}$ (the length of the error bars represents the standard deviation of the estimated mean bias for all the simulate datasets.). The horizontal dotted line ($\cdot\,\cdot\,\cdot\,\cdot\,\cdot$) represents the corresponding estimates from the full model fitted with MCMC. The horizontal dashed line ($------$) indicates the value at which the results are unbiased.}
\end{figure}

Figure \ref{lgssm} presents the results of the median bias and MCSE of model parameters and latent states estimated from the fitted linear Gaussian SSMs using both MCMC and the proposed Monte Carlo algorithm. The results from the full models fitted using MCMC are displayed as the dashed horizontal line (the results from the full models fitted with pMCMC are displayed in Supplementary Information Two Figure 1). Figure \ref{fig-psifs} shows the correlation and bias in growth rate estimated from the fitted dynamic occupancy model.

We observed from both simulation studies that the distribution of the model parameters was comparable between the updated models (BUMC and AUMC; Figure \ref{lgssm}, Table S3-1). The estimates of the model parameters were not very different between the reduced and updated models. However, as $t$ (the number of years used to fit the reduced model) approached $T$ (the number of years used to fit the updated models), the model parameter estimates from the updated and full models were very similar, with the bias of the estimates closer to $0$. Moreover, the MCSE for some parameters from the updated models was smaller than those from the full model (Figure \ref{lgssm}). This suggests that the performance of the proposed algorithm in estimating the true parameter values depends on the estimates from the reduced model (this is what we expect given the proposal distribution specified for the Metropolis-Hastings algorithm, see section \ref{updating-the-metroplis-hasting-approach}).

For all the updated and full linear Gaussian SSMs fitted, the true latent state distribution was estimated (with median bias over the $90$ replicated simulations around $0$; Figure \ref{lgssm}) accurately. For the dynamic occupancy model fitted in simulation two, we observed that the correlation in the realised occupancy probability from the updated models
was similar for both updated and full models, irrespective of the time used to fit the reduced model $t$ and the number of particles ($M$). These observations indicate that the proposed Monte Carlo algorithm can also capture the latent state distribution (and trends of the derived quantities of the latent states) equally well compared with the full model.

\begin{figure}[htbp!]
\centering 
\includegraphics{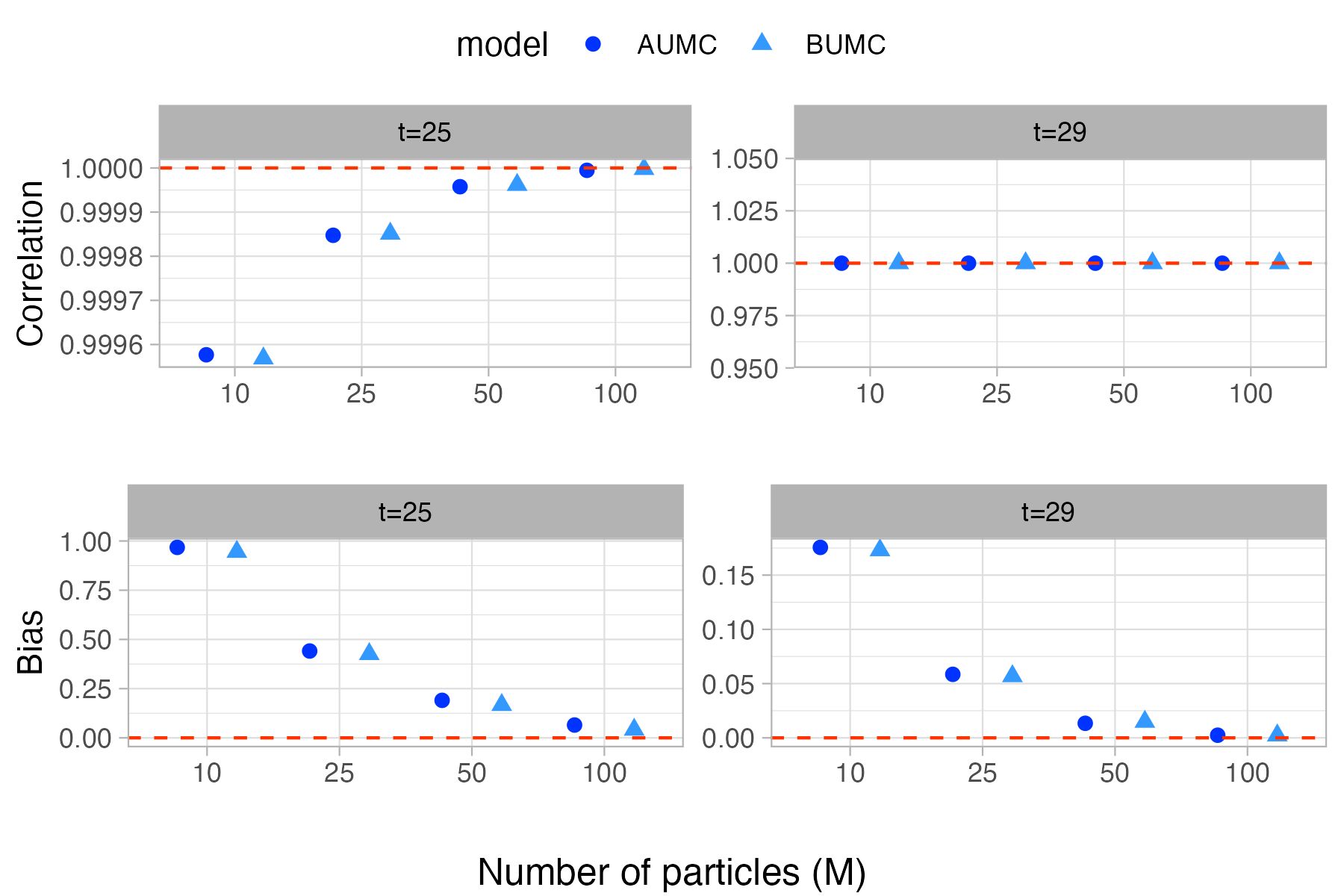}
\caption{\label{fig-psifs} First row: Correlation between the estimated realised occupancy probability and true value from simulated data for years $t+1$ to $30$. Second row: Bias in the realised occupancy probability (expressed as a percentage) is estimated as the difference between the estimated realised occupancy from the fitted models and the true values from the simulated data. We used two different years to fit the reduced model $t = 25$ and $t = 29$, and then we fit the updated models using four values of $M$: $10$, $25$, $50$ and $100$.}
\end{figure}

\subsubsection{Efficiency of proposed Monte Carlo algorithm}
For the dynamic occupancy model in section \ref{simulation-study-two}, the effective sample size from the proposed Monte Carlo algorithm was lower than that from MCMC, but the proposed algorithm was more efficient in generating effective samples per second (Supplementary Information Two Figures 4 and 5). The efficiency of the proposed Monte Carlo algorithm comes from the computational time it used in fitting the dynamic occupancy model (Table \ref{tab-computationaltime}). 

We observed that the computational time for the updated models (AUMC and BUMC) increased by $O(n)$ (i.e. increased linearly) with the number of particles ($M$) and the number of years used to fit the reduced model ($t$; Table \ref{tab-computationaltime}). However, the increase in the computational time $t$ is relatively smaller than that of the number of particles $M$. This suggests that much effort should be put into choosing an optimal number of particles to obtain much efficiency from the proposed Monte Carlo algorithm.

Additionally, the computational times from the updated dynamic occupancy models were similar for both particle filter algorithms: bootstrap and auxiliary particle filter (Table \ref{tab-computationaltime}). There is however a slight increase in the computational times from the auxiliary particle filters, which is due to its look-ahead step \citep[i.e. at each time step after $t$, we first sample $M$ particles using the weights from the previous time points to obtain a rough estimate of the likelihood of the current data before the sampled particles are propagated in time and re-weighted using another weight][]{michaud2021sequential, pitt1999filtering}.

\begin{table}[ht!]
\centering
\begin{tabular}{ |p{1.5cm}|p{1cm} p{1cm} p{1cm} p{1cm}|p{1cm} p{1cm} p{1cm} p{1cm}|}
\hline
\multicolumn{1}{|c|}{} & \multicolumn{4}{c|}{t = 29} & \multicolumn{4}{c|}{t = 25} \\
\hline
Model & M=10 & M=25 &M=50 & M=100 & M=10 & M=25 & M=50 & M=100\\
\hline
BMC & \multicolumn{4}{c|}{25} & \multicolumn{4}{c|}{25} \\
BUMC & 2.67 & 4.45 & 7.22 & 12.25 & 2.93 & 5.36 & 9.47 & 17.11 \\
AUMC & 2.93 & 4.88 & 8.26 &  13.37 & 3.57 & 6.90 & 11.89 & 21.73 \\
\hline
\end{tabular}
\caption{\label{tab-computationaltime} Computational times (in minutes) of the various algorithms used to fit the dynamic occupancy models. The models fitted are summarised in Table \ref{table:modelSummary}. Two different time steps ($t = 29,25$) were used to fit the reduced model, and four number of particles ($M = 10, 25, 50, 100$) were chosen to fit the updated models.}
\end{table}

\subsubsection{Convergence of MCMC chains}
The convergence trace plots of model parameters estimated from the baseline and updated models are presented in Supplementary Information Two, Figures 2 and 3 respectively. The full and updated models fitted using the pMCMC approach with an auxiliary particle filter showed convergence issues, with the other approaches converging for all model parameters. The poor convergence of the pMCMC approach using an auxiliary particle filter likely led to the high estimates of bias and MCSE of model parameters for this approach.

\subsection{Case Study}
\subsubsection{Case study 1}\label{results_casestudy1}

We present the estimated interannual growth rate and population size from the demographic SSM fitted using the MCMC (full model; BMC) and the proposed Monte Carlo algorithm (updated model; AUMC and BUMC) in Table \ref{tab-populationdemographics}. The interannual growth rates from both full and updated models are similar, even though the estimated population size was higher in the full models than the updated models. The computational time for both updated population demographic models (AUMC and BUMC) takes about a third of the time to fit the model again (Table \ref{tab-populationdemographics}).

\begin{table}[h!]
\centering
\begin{tabular}{ |p{3cm}|p{2.5cm}|p{1.5cm}|p{1.5cm}|p{1.5cm}|p{1.5cm}|  }
\hline
\multicolumn{1}{|c|}{} & \multicolumn{1}{|c|}{} &  \multicolumn{2}{|c|}{Interannual rate} & \multicolumn{2}{|c|}{Population size} \\

Model & Computational time (minutes) & Median & SD & Median & SD \\
\hline
BMC & 16.35 & 1.011 & 0.031 & 1883 & 148.53  \\
BUMC & 5.02 & 1.012 & 0.0703 & 1815 & 138.43  \\
AUMC & 5.66 & 1.017 & 0.066 & 1816 & 138.43 \\
\hline
\end{tabular}
\caption{\label{tab-populationdemographics} Computational times (in minutes), posterior mean and standard deviation of the interannual rate and population size in the year 2016 estimated from the population demographic model fitted to the Swiss Crested Tits. The full model (BMC) was fitted to the observed data from 1996 - 2016 using MCMC and the two updated models (AUMC and BUMC) were fitted using the proposed Monte Carlo algorithm.}
\end{table}

\subsubsection{Case study 2}

Table \ref{tab-spartaOccupancy} shows the posterior median (with standard deviation in parenthesis) realised occupancy and growth rate of the Yellow Meadow ant in $2021$. The occupancy models fitted with the proposed algorithm (AUMC and BUMC) and the full model fitted with MCMC produced comparable results of realised occupancy and growth rate.  Furthermore, the computational time to re-fit the occupancy model again using MCMC via the R-package \textit{nimble} is reduced significantly when the proposed algorithm is used to update the already fitted occupancy model. This observation supports an earlier assertion made in section \ref{results_casestudy1}.

\begin{table}[h!]
\centering
\begin{tabular}{ |p{2.5cm}|p{2.5cm}|p{3cm}|p{1.5cm}|  }
\hline
%\multicolumn{1}{|c|}{} & \multicolumn{1}{|c|}{} &  \multicolumn{1}{|c|}{Realised Occupancy ($\Psi$)} & \multicolumn{1}{|c|}{Growth rate} \\

Model & Computational time (minutes) & Realised Occupancy ($\Psi$) \\
\hline
BMC (sparta) & 40 & 0.44 (0.047)  \\
BMC (nimble) & 281 & 0.44 (0.059) \\
BUMC & 48.26 & 0.50 (0.046)   \\
AUMC & 59.93 & 0.40 (0.095)   \\
\hline
\end{tabular}
\caption{\label{tab-spartaOccupancy} Computational times (in minutes), realised occupancy probability (posterior median with standard deviation in parenthesis) in the year 2020 estimated from the occupancy model fitted to the ant dataset. The full model was fitted to the observed data with the R-package \textit{sparta} (BMC(sparta)) and \textit{nimble} (BMC(nimble)) and the two updated models (AUMC and BUMC) were fitted using the proposed Monte Carlo algorithm.}
\end{table}

\hypertarget{discussion}{%
\section{Discussion}\label{discussions}}

In this study, we have proposed a computationally efficient Monte Carlo algorithm to update an already-fitted state-space model. The algorithm capitalises on the strength of importance sampling to update the posterior distribution of the latent state and MCMC to update the posterior distribution of the model parameters. We use simulation and case studies to assess how well the proposed algorithm estimates the distribution of the model parameters, the Monte Carlo standard error of the samples obtained from the algorithms, how many effective samples the algorithm samples per second, and the computational time it takes to fit SSMs using our proposed algorithm instead of re-fitting models again using MCMC. We implement this proposed Monte Carlo algorithm by copying and editing the particle filter and Metroplis-Hastings algorithms implemented in the R-package \textit{nimbleSMC} \citep{michaud2021sequential}. We observed that the proposed algorithm estimated the model parameters and derived quantities of latent states as well as re-fitting the models again, but with a relatively shorter computational time.

The motivation for this work came from previous work fitting occupancy models with \textit{sparta} R-package \citep{august} to biological records \citep{outhwaite2019annual}. Among the objectives of fitting these occupancy models is to estimate the trend of occupied sites over time \citep{outhwaite2019annual}, which is used to provide regular (e.g. annual) reports of the state of nature. These models take a long time to run, and re-running the models every time new data becomes available is impractical (the ant dataset used in the case study two is among the smallest of 20 taxonomic groups in \citeauthor{outhwaite2019annual}). There was a need to provide an efficient approach to obtain the yearly estimates of the trends of occupied sites once the yearly data became available.

 In modelling ecological time-series biodiversity data, temporal trends of derived quantities that describe species distribution, such as growth rate, population size and realised occupancy, are of interest \citep{royle2007bayesian, outhwaite2019annual}, as well as the drivers (e.g. covariate effects) \citep{kery2020applied}. Fitting SSMs with our proposed Monte Carlo algorithm captures the latent state (and functions of the latent state such as realised occupancy probability; \citealp{royle2007bayesian} and interannual growth rate; \citealp{kery2020applied}) and model parameters distribution as well as re-fitting the SSM again via MCMC, as revealed from the simulation and case studies (Figures \ref{lgssm} and \ref{fig-psifs}; Tables \ref{tab-populationdemographics} and \ref{tab-spartaOccupancy}). This is unsurprising as the bootstrap and auxiliary particle filters provide unbiased estimates of the marginal likelihood \citep{michaud2021sequential, andrieu2010particle}. However, the results from the proposed algorithm are obtained in a fraction of the time it takes to re-fit the SSMs again using MCMC, especially when updating the already-fitted SSM annually (Tables \ref{tab-computationaltime}, \ref{tab-populationdemographics} and \ref{tab-spartaOccupancy}). This indicates that the algorithm we propose in this study provides a feasible alternative to re-run these computationally expensive models within a relatively shorter time. 

 The proposed algorithm has some properties we explored in this study. The proposed algorithm is a revised version of the particle MCMC algorithm \citep{andrieu2010particle}. We have assumed that we have already fitted the SSM using MCMC to the observed data up to time $t$ and have saved the posterior distribution of the latent state and model parameters. We then copy this posterior distribution into the particle algorithm that updates the already-fitted SSM. The simulation study revealed that the performance of the proposed algorithm in capturing the actual model parameter values depends on how well the posterior distribution from the already-fitted model is estimated (Figure \ref{lgssm}), as new values are proposed from a distribution that depends on the posterior distribution from the already-fitted model (section \ref{updating-the-metroplis-hasting-approach} and Supplementary Information One Algorithm 4). This suggests that careful attention should be paid to the information stored in the fitted models. Ideally, the posterior distribution of all of the model parameters (especially stochastic variables) and the latent state distribution at time $t$ should be stored (see equation \ref{framework}). In practice, however, this would require more infrastructure to maintain the stored information over the long term. Consequently, fewer samples are saved for future use. In such cases, more samples can be drawn from the posterior distribution through different importance sampling techniques \citep[such as in ][]{bhattacharya2007importance, cruz2022iterative}, and the drawn samples can be used to update the existing model.

SSMs can be computationally expensive (or slower), especially for high dimensional model parameters and latent state space \citep{newman2023state}. With an increasing volume of datasets generated yearly (such as those from novel technologies \citep{hartig2024novel}), the run times of the algorithms can become a bottleneck for biodiversity analyses \citep{boyd2023operational}. The computational time of the proposed algorithm, just like the pMCMC algorithm it revises, depends on the number of particles \citep[$M$; ][]{michaud2021sequential, andrieu2010particle} and the number of years used to fit the reduced model $t$. From the simulation and case studies, we observed that the computational time increases linearly (with factor $O(n)$) as the number of particles $M$ increased (Tables \ref{tab-computationaltime}, \ref{tab-populationdemographics} and \ref{tab-spartaOccupancy}). The studies further revealed that the computational time of the proposed algorithm increases with the number of years ahead we aim to predict or make inferences on, from the time $t$ we fitted the existing SSM. However, the number of years cannot be controlled in reality, since the existing models would need to be updated with these new observed data. Additionally, the type of particle filter fitted had little effect on the computational time. This hinted that an optimal choice of the number of particles $M$ is crucial to attain the computational advantage our proposed model presents.

Choosing an optimal value of $M$ is not a cast-in-stone one. Studies exploring the effect of $M$ on the validity of the pMCMC algorithm (which our proposed algorithm modifies) have shown that the algorithm is valid for small values of $M$, but there is a risk of the algorithm getting stuck due to high variance in the model likelihood approximation with the particle filters \citep{andrieu2010particle, michaud2021sequential, sherlock2015efficiency,doucet2015efficient}. Further studies show that the quality of the likelihood approximation affects how the chain mixes and the efficiency of the pMCMC \citep{andrieu2015convergence, andrieu2016establishing, fearnhead2018particle}. Therefore, a good choice of $M$ could be made to generate efficient samples at a low computation cost by updating the existing models with fewer particles and iteratively increasing it as the efficiency of the sample is monitored \citep{michaud2021sequential, fearnhead2018particle}. Also, good initial values need to be specified for the model parameters \citep[such as using maximum \textit{a posteriori} estimates][]{michaud2021sequential, bernton2017inference}. In the case study, we used the posterior mean from the reduced model as initial values for the model parameters. The particles in the proposed Monte Carlo at each iteration are drawn independently of each other. Therefore, the algorithm can be parallelised to improve the computational time for a given number of particles ($M$). 

Another factor that contributes to the efficiency of the pMCMC algorithm is the sampler used to generate samples of the model parameters. The R-package \textit{nimbleSMC}, which we copied and edited to implement our proposed algorithm, jointly samples all the model parameters from a random walk block sampler \citep{michaud2021sequential}. This sampler is very efficient for smaller dimensions of the model parameters \citep{tibbits2014automated}. However, for higher dimensional model parameters (as in the dynamic occupancy model in section \ref{simulation-study-one}, population demographic models in section \ref{case-study-one} and the occupancy model in section \ref{case-study-two}), the pMCMC exhibits 'stickiness' for model parameters (where the model parameters get stuck at some proposed values) when jointly sampled with the default sampler implemented in \textit{nimbleSMC}. We implement a modified block sampler that first samples the hyperparameters, and the rest of the model parameters are sampled afterwards (see Supplementary Information One Algorithm 4). This proposed sampling approach may not be the best to handle such high dimensional model parameters for pMCMC; however, it sets the course for further developments of sampling algorithms for the R-package \textit{nimbleSMC}. Further studies can explore Hamiltonian Monte Carlo \citep{neal2011mcmc} or slice samplers \citep{neal2003slice} as alternatives to the Metropolis-Hastings block sampler. 

We also observed from the simulation and case studies that the complexity of the SSM fitted and the implementation software could have an impact on the computational efficiency of our proposed algorithm. The occupancy model fitted in section \ref{case-study-two} has sparse latent state distribution, and the R-package \textit{sparta} - using JAGS software - fits this model within $40 minutes$. The same model, when fitted with the R-package \textit{sparta} - using NIMBLE software (which is currently being added to the implementation framework of \textit{sparta}) - takes over four hours (with the same MCMC configurations). Due to the differences in the computational times from the different software (NIMBLE and JAGS), we only use the posterior samples from the model fitted with \textit{sparta} to fit the updated models but do not make any comparison between the full model fitted using \textit{sparta} and the proposed Monte Carlo algorithm. These computation differences in efficiency from the different software have been explored in other studies \citep{beraha2021jags, li2018fitting}. We acknowledge however that these differences in computational times could be problem-specific, and be dependent on the choice of MCMC samplers \citep{ponisio2020one, turek2017automated}.

To draw the curtains on the properties of the proposed Monte Carlo algorithm in this study, particle degeneracy or inefficiency \citep{knape2012fitting, newman2023state} - which occurs when the particles drawn are not representative of the observed data \citep{knape2012fitting} - could affect the performance of the proposed algorithm. Particle degeneracy becomes a problem when the observed data are very informative about the hidden states (such as those in the population demographic and occupancy models). To circumvent this issue in the occupancy models, for example, we only simulate the latent state whose corresponding observation was absent from the prior distribution. This way, we avoid having the true latent state absent when the corresponding observation is present (i.e. $P(y= 1|z=0) = \infty$).

The algorithm proposed in this study can also be applied to non-time series models. For example, particle filters have been used to update the static parameters from generalised linear and mixture models when newly observed data is available using the iterated batch importance sampling algorithm \citep{chopin2002sequential}. Further studies can also explore using this iterated batch importance sampling algorithm to simulate the model parameters instead of the Metropolis-Hastings algorithm. 

The proposed algorithm is a flexible and easy-to-implement Monte Carlo algorithm to update already-fitted SSM with new observed data. This makes it possible to integrate into workflows that model biodiversity data generated sequentially \citep[such as those developed in][]{boyd2023operational, outhwaite2019annual}. The proposed algorithm is also computationally efficient, taking a relatively short time to run. This makes it feasible to handle models aimed at capturing in real time the trend in latent states or model parameters (or functions of both) using animal movement and population dynamics, mark-recapture datasets. 

\section*{Data availability}\label{data-availability}
All the code used for the analysis and plotting of figures is available on the GitHub repository and archived on Zenodo with DOI: 10.5281/zenodo.10463704. In addition, the changes in the \textit{nimbleSMC} R-package described in Supplementary Information One are hosted in the R-package \textit{nimMCMCSMCupdates} \citep{nimMCMCSMCupdates}.

\section*{Author contribution}\label{data-availability}
KPA and RBO were involved in the methodology development, with NI and RBO contributing to the idea development. KPA further did the analysis and wrote the R-package and first draft of the manuscripts. All authors were involved in writing the manuscript.

\section*{Conflict of Interest}
The authors declare no conflict of interest.

\section*{Acknowledgement}
This study is part of the Transforming Citizen Science for Biodiversity project funded by the Digital Transformation initiative of the Norwegian University of Science and Technology. We also acknowledge the contributions of Diana Bowler in developing the methodology in this study. NJBI was supported by the Natural Environment Research Council award number NE/R016429/1 as part of the UK-SCAPE programme delivering National Capability. We also acknowledge the contribution of Daina Bowler in developing the methods presented in this study.

\bibliographystyle{rss}
\bibliography{references}
\end{document}